# Brownian Motion of Arbitrarily Shaped Particles in Two-Dimensions

*Ayan Chakrabarty[1†], Andrew Konya[1†], Feng Wang[1], Jonathan V. Selinger[1*], Kai Sun[2], Qi-Huo Wei[1*]*

1Liquid Crystal Institute and Department of Chemical Physics, Kent State University, Kent, OH 44242, USA

2 Department of Materials Science and Engineering, University of Michigan, Ann Arbor, MI 48109, USA

**KEYWORDS:** Brownian motion, asymmetric colloids, single particle tracking, translation-rotation coupling

**ABSTRACT:** Here we implement microfabricated boomerang particles with unequal arm lengths as a model for non-symmetry particles and study their Brownian motion in a quasi-two dimensional geometry by using high precision single particle motion tracking. We show that due to the coupling between translation and rotation, the mean squared displacements of a single asymmetric boomerang particle exhibit a non-linear crossover from short time faster to long time slower diffusion, and the mean displacements for fixed initial orientation are non-zero and saturate out at long time. The measured anisotropic diffusion coefficients versus the tracking point position indicate that there exists one unique point, i.e. the center of hydrodynamic stress (CoH), at which all coupled diffusion coefficients vanish. This implies that in contrast to motion in 3D where the CoH only exists for high symmetry particles, the CoH always exists for Brownian motion in 2D. We develop an analytical model based on Langevin theory to explain the experimental results and show that among the 6 anisotropic diffusion coefficients only 5 are independent because the translation-translation coupling originates from the translation-rotation coupling. Finally we classify the behavior of 2D Brownian motion of arbitrarily shaped particles into four groups based on the particle shape symmetry group.

## ■ INTRODUCTION

Brownian motion as the essential process of diffusion and mass transport is critical to many physical, chemical and biological processes.[1-3] More than one century after its first theoretical explanation by Einstein,[4] intriguing physics and applications of Brownian motion are still emerging from various investigations.[5-14] For example, by using high precision optical trapping and motion tracking techniques, Franosch *et al* observed the resonances originating from hydrodynamic memory and the long-sought colored spectrum of the thermal force in Brownian motion,[5] and Raizen and coworkers observed for the first time the ballistic motion at short time scale and the transition from short-time ballistic to long-time diffusive Brownian motion.[6-7] Turiv *et al* showed that Brownian motion in nematic liquid crystals is not only anisotropic but also anomalous and that the diffusive regimes are strongly influenced by the deformation and fluctuations of the molecular orientational order.[9]

Colloids of exotic anisotropic shapes are being explored as building blocks for self-assembly of novel materials[15-21] and efficient carriers for drug delivery.[22] As the symmetry of particle shapes has significant influence on hydrodynamic properties,[23-29] profound understanding of the influence of particle shape on Brownian motion is important to a variety of problems such as electrophoresis,[30] sedimentation,[31-33] micro-swimmers,[34-37] molecular/particle sorting[38-40] and self-assembly.[41-43] While hydrodynamic theory has been established by Brenner and others for Brownian motion of arbitrarily shaped colloidal particles,[23-29] particles of simple shapes such as spheres,[8-9, 44-45] rods[46] and ellipsoids[47-48] are mostly studied in experiments. It was only recently that experimental studies on Brownian motion of more complex shaped particles started to emerge.[49-55] In two dimensions (2D), circular disks represent the highest symmetry $D_\infty$, and their diffusion can be described by one translational and one rotational diffusion coefficient. Elliptical disks with $D_2$ symmetry can be described by two translational and one rotational diffusion coefficient. Studies by Han *et al* demonstrated that while the mean square displacements (MSDs) of ellipsoids in 2D remain Fickian all time, their displacement probability distribution functions are non-Gaussian at short time due to the memory of anisotropic diffusion.[47] The translational and rotational motions for these ellipsoids and other high symmetry particles are decoupled because their center of mass (CoM) normally used for motion tracking is coincident with the symmetry center.

For particles without any symmetry center, the selection of tracking points (TPs) becomes non-trivial because translation and rotation are normally coupled. Recently, we showed that Brownian motion of symmetric boomerang particles is remarkably different from that of spheres and ellipsoids.[54] Boomerangs with two equal arms possess the $D_1$ symmetry, the lowest symmetry in the 2D dihedral point group. We showed that the mean displacements (MD) are biased towards a unique point, i.e. the center of hydrodynamic stress (CoH), where the translation and rotation are decoupled. As a result, the MSDs exhibit a nonlinear crossover from short time faster diffusion to long time slower diffusion. Their diffusion process in 2D can be described by four independent diffusion coefficients: two translational, one rotational and one coupling diffusion coefficient. However, the Brownian motion of arbitrarily shaped particles with no shape symmetry has never been unexplored, which is a missing element to the complete physical picture of shape effects in 2D Brownian motion.

In this paper, we employ asymmetric boomerang particles confined between two parallel glass plates as a model system for nonsymmetrical rigid bodies in 2D and study their Brownian motion by using high precision single particle tracking. We show that for fixed initial orientation, the MDs are nonzero and the MSDs exhibit two distinct linear regimes with different diffusion coefficients and a non-linear crossover regime. For a body frame fixed on the particle, the translational motions along two body frame axes are normally coupled with rotation. In particular, translational motions along two body frame axes are coupled with each other, which is in contrast to symmetric boomerangs. This translation-rotation coupling necessitates two translational, one rotational and three coupling diffusion coefficients for a complete description of their motion. By measuring the variations of diffusion coefficients with the TPs, we show that there exists a unique point where all three coupling diffusion coefficients vanish. To elucidate the experimental observations, we develop a model based on Langevin theory and obtain excellent agreement with the experimental data. The model further demonstrates that among these 6 diffusion coefficients, only 5 are independent. At last, we classify the behavior of 2D Brownian motion of arbitrarily shaped particles into four groups based on the symmetry point group of their shapes.

■ **EXPERIMENTAL METHODS**

To fabricate the asymmetric boomerang particles, we spin-coated a 17nm film of soluble polymer (Omnicoat, Microchem) on clean Si wafers as the sacrificial layer and then a 500nm film of UV-curable photoresist (SU8, MicroChem). The photoresist films were patterned by using an autostepper projection photolithography system through standard lithography processes. After developments of the patterned boomerang particles, the Si wafers were submerged in a solvent (PG remover, MicroChem) to disperse the particles and ultrasonic shaking is used to expedite the process. The solvent is replaced by deionized water through centrifugation. A low concentration of sodium dodecyl sulfate (1mM) is added into the particle dispersion for stabilization.

An exemplary SEM picture of the fabricated particles is shown in Figure 1a. The particle used for this study has a long arm of 2.25 μm in length and a short arm of 1.3 μm in length, and a 90° apex angle. The arm width is 0.7 μm. Sample cells composed of two parallel glass slides were prepared with ~1.7 μm cell thickness, which is thin enough to restrict the motion of the particles in quasi-two dimensions. Videos of individual isolated asymmetric boomerangs were recorded at the time step $\tau = 0.05$ s with an electron-multiplying charge coupled device (EMCCD, Andor Technology). Each video contains 3000 image frames. 201 videos for the same particle with the same time intervals were recorded and used for the study.

To determine the position and orientation of asymmetric boomerang particles, we employed a high precision tracking algorithm developed by us.[55] The algorithm finds the central axes of boomerang arms by fitting the intensity profile across the arms with Gaussian point spread functions, and then determine the particle position and orientation by finding the cross point of two central axes and the orientation angle of the bisector line. As detailed in the previous publication, the precisions of our optical microscopy and tracking algorithms are very high, with ±13nm for position and ±0.004 rad for orientation.[55] This high precision tracking is necessary for the measurements of non-zero MDs for fixed initial angle. Based on the fact that the displacements at two sequential time intervals are uncorrelated, we merged the 201 videos into single motion trajectory. The process of merging two videos is to translate and rotate the coordinates in the second video so that the first frame in the second video matches the last frame of the first video.[55]

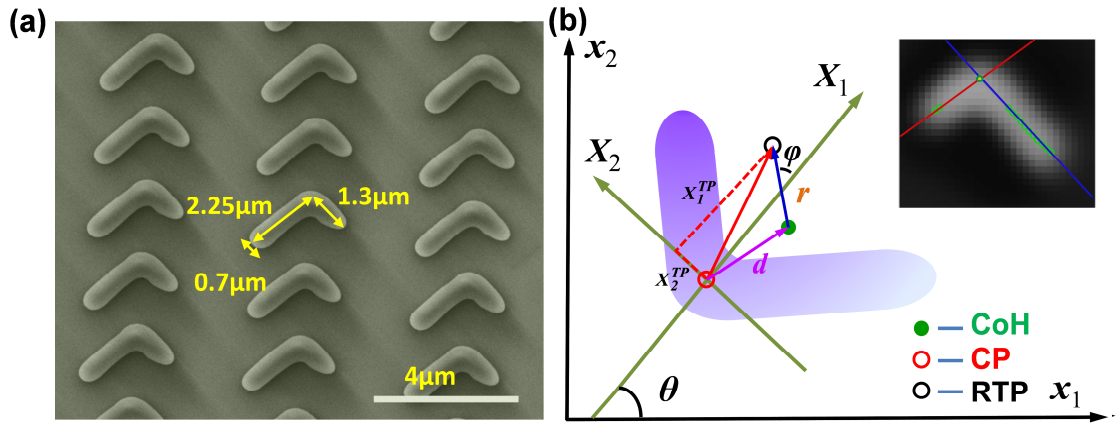

**Figure 1. Boomerang particles and the coordinate systems:** (a) SEM image of the asymmetric boomerang particles fabricated on a silicon wafer. (b) Schematics of the lab frame ($x_1$-$x_2$) and the body frame ($X_1$-$X_2$) coordinate systems. The red and black open circles represent the cross point (CP) of two arms and a random tracking point (RTP); the green dot represents the CoH. The coordinates of the TP are denoted as $\left(X_1^{TP}, X_2^{TP}\right)$, the vector from the CoH to the TP is denoted as $\boldsymbol{r}$, the vector from the CP to the CoH is denoted as $\boldsymbol{d}$, and the angle between $\boldsymbol{r}$ and $X_1$ axis is denoted as φ. The inset shows a typical processed optical microscopic image of a boomerang particle.

■ **Experimental Results**

The asymmetric boomerang particles used in the experiments are made of photo curable polymers (SU8) with two different arm lengths and a $90^0$ apex angle (Figure 1a). Aqueous suspensions of the boomerang particles are confined between two parallel glass substrates with 1.7 μm cell spacing. A total of 201 videos were recorded for the Brownian motion of a single isolated boomerang particle. Limited by computer memory, each video contains 3000 image frames. These videos were processed and then combined to construct a single motion trajectory by using the high precision image processing algorithm developed by us.[55] The cross point of the central axes of two arms is used to represent the particle position and the bisector line of the apex angle is used to represent the particle orientation. The lab frame and body frame coordinate systems are schematically shown in Figure 1b. To note, the CoM in this case lies outside the body and is not a convenient point for motion tracking because minute out-of-plane rotation deteriorates its location accuracy in image processing.

Figure 2a presents the measured MSDs in the lab frame. The agreements between the MSDs along $x_1$ and $x_2$ axes suggest that the Brownian motion is isotropic overall. The MSDs grow linearly with time only at short and long time (Figure 2a). Best linear fittings yield a short time diffusion coefficient $D^{ST} = 0.0375\, \mu m^2$/s for time $t < 10$ s and a long time diffusion coefficient $D^{LT} = 0.0225\, \mu m^2$/s for $t > 100$ s. The existence of a sub-diffusive crossover regime between 10 s and 100 s is distinctively different from high symmetry particles such as spheres and ellipsoids where the MSDs always grow linearly with time. The measured MSDs of the angular orientation grow linearly with time with the angular diffusion coefficient $D_\theta$ = 0.034 rad²/s (Figure 2b).

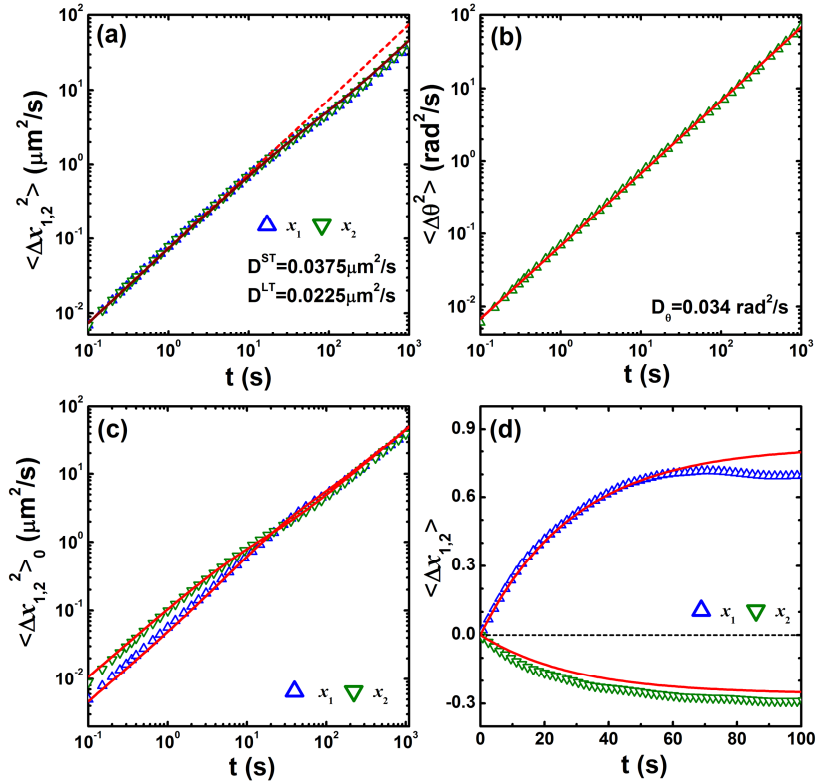

**Figure 2. Lab Frame Results:** (a) MSDs vs. $t$. The red dash line is a linear fit for $t < 10$s and the dark brown solid line is a plot of Eq. 4, with $\overline{D}^{LT} = 0.0225\, \mu m^2/s$, $\overline{D}^{ST} = 0.0375\, \mu m^2/s$. (b) MSDs of the orientational angle $\theta$ vs. $t$. The red line is the linear fit with $D_\theta = 0.034$ rad$^2$/s. (c) MSDs with $\theta_0 = 0$ vs. $t$. Red lines are theoretical curves of Eq. 3. (d) MDs along the $x_1$ and $x_2$ directions for $\theta_0 = 0$. The red lines are theoretical curves using Eq. 2.

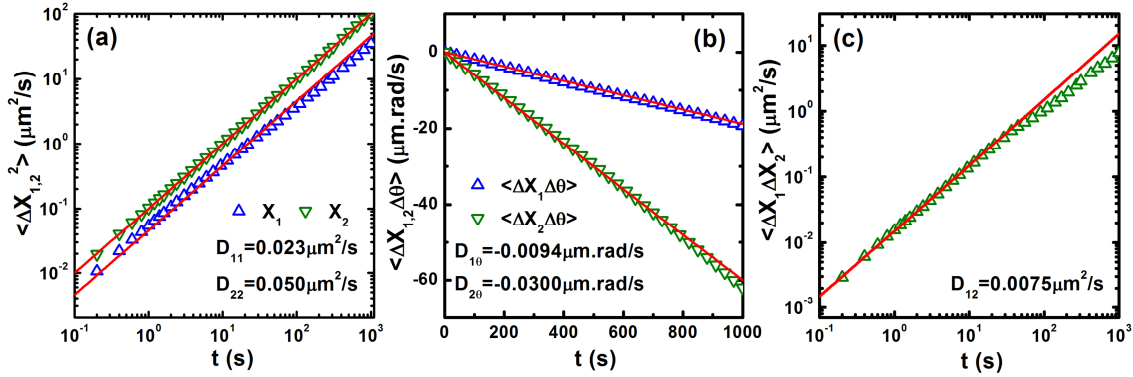

**Figure 3. Body Frame Results:** (a) MSDs along the $X_1$ and $X_2$ directions vs. $t$. The red solid lines are linear fits with $D_{11}$ = 0.023 μm²/s, $D_{22}$ = 0.050 μm²/s. (b) Translation-rotation correlations vs. $t$ along the $X_1$ and $X_2$ directions. The red solid lines are linear fits with $D_{1\theta}$ = -0.0094 μm.rad/s, $D_{2\theta}$ = -0.0300 μm.rad/s (c) Correlations between displacements along the $X_1$ and $X_2$ direction. The red line is a linear fit with $D_{12}$ = 0.0075 μm²/s.

Due to the asymmetry of the boomerang shape, the Brownian diffusion is expected to be anisotropic in short time. This behavior can be observed in the MSDs calculated with a fixed initial orientation ($\theta_0$) of the particle. As shown in Figure 2c, for $\theta_0$=0, the diffusion along $x_2$ axis at short time ($t$ <10s) is faster than along $x_1$ axis, while at long time ($t$ >100s), the diffusion coefficients along two axes become the same. In comparison to ellipsoidal particles where the MSDs along two axes converge at long times,[47] the MSDs along $x_1$ are larger than those along $x_2$, similar to our previous observations in the case of symmetric boomerang particles.

We also calculated the MDs for $\theta_0$ = 0 and observe that the MDs are positive along the $x_1$ axis but negative along the $x_2$ axis (Figure 2d) and saturate out at long time in both cases. To ensure that this biased MD is not due to any directional drift in the system, we calculated the MDs averaged over different initial orientations and verified that they are zero over all time. This non-zero MDs for fixed initial orientation is in sharp contrast to highly symmetric particles where the MDs are always zero, and is different from the symmetric boomerang particles where the MDs are non-zero only along the symmetry axis.[54]

To measure all elements of the diffusion coefficient tensor, we convert the displacements in the lab frame into those in the body frame through a rotation transformation: $\Delta X_i(t_n) = R_{ij}(\theta_n)\Delta x_j(t_n)$, where $i, j = 1$ or $2$, $\Delta x_i(t_n) = x_i(t_{n+1}) - x_i(t_n)$ is the displacement in the lab frame between two consecutive time steps and $R_{ij}(\theta_n)$ is the rotation transformation matrix. As shown for the symmetric boomerangs, the choice of $\theta_n$ is not trivial when the TP is not coincident with the CoH, and $\theta_n = [\theta(t_n) + \theta(t_{n+1})]/2$ needs to be used to obtain proper body frame results.[54] We follow this notation and construct the body frame trajectory using $X_i(t_n) = \sum_{k=1}^{n} \Delta X_i(t_n)$.

The MSDs in the body frame grow linearly with time, yielding two anisotropic diffusion coefficients along the $X_1$ and $X_2$ directions, $D_{11} = 0.025 \mu m^2/s$ and $D_{22} = 0.05 \mu m^2/s$ (Figure 3a). As expected from the asymmetric particle shape, translation and rotation are coupled; $\langle \Delta X_1 \Delta \theta \rangle$ and $\langle \Delta X_2 \Delta \theta \rangle$ decrease linearly with time, implying negative coupled diffusion coefficients (Figure 3b). The coupled diffusion coefficients obtained from linear fittings are, $D_{1\theta} = -0.0095$ μm•rad/s and $D_{1\theta} = -0.0305$ μm•rad/s. In contrast to symmetric boomerangs where translations along two axial directions are decoupled, the translations of the asymmetric boomerangs in the $X_1$ and $X_2$ directions are coupled (Figure 3c), and the measured coupled diffusion coefficient is $D_{12} = 0.0075$ μm²/s.

To determine if the translation and rotation can be decoupled, we reconstructed the motion trajectories by using different TPs and then recalculated the diffusion coefficients. The resulted anisotropic diffusion coefficients are presented in Figure 4 as functions of the TP coordinates $(X_1^{TP}, X_2^{TP})$. We can see that $D_{11}$ and $D_{1\theta}$ remain constant for different $X_1^{TP}$ but vary significantly with $X_2^{TP}$, while the coupled translation diffusion coefficient $D_{12}$ varies with the TP coordinates in both $X_1$ and $X_2$ directions. In addition, we can observe that at $X_2^{TP} = 0.28\,\mu m$ both $D_{1\theta}$ and $D_{12}$ approach zero while $D_{11}$ reaches its minimum, and that at $X_1^{TP} = 0.88\,\mu m$, both $D_{2\theta}$ and $D_{12}$ approach zero while $D_{22}$ reaches its minimum. In another word, all translation-rotation and translation-translation coupling disappear when the point fixed to the particle at (0.88μm, 0.28 μm) is used for motion tracking. This indicates that the asymmetric boomerang particle has its CoH located at (0.88μm, 0.28 μm).

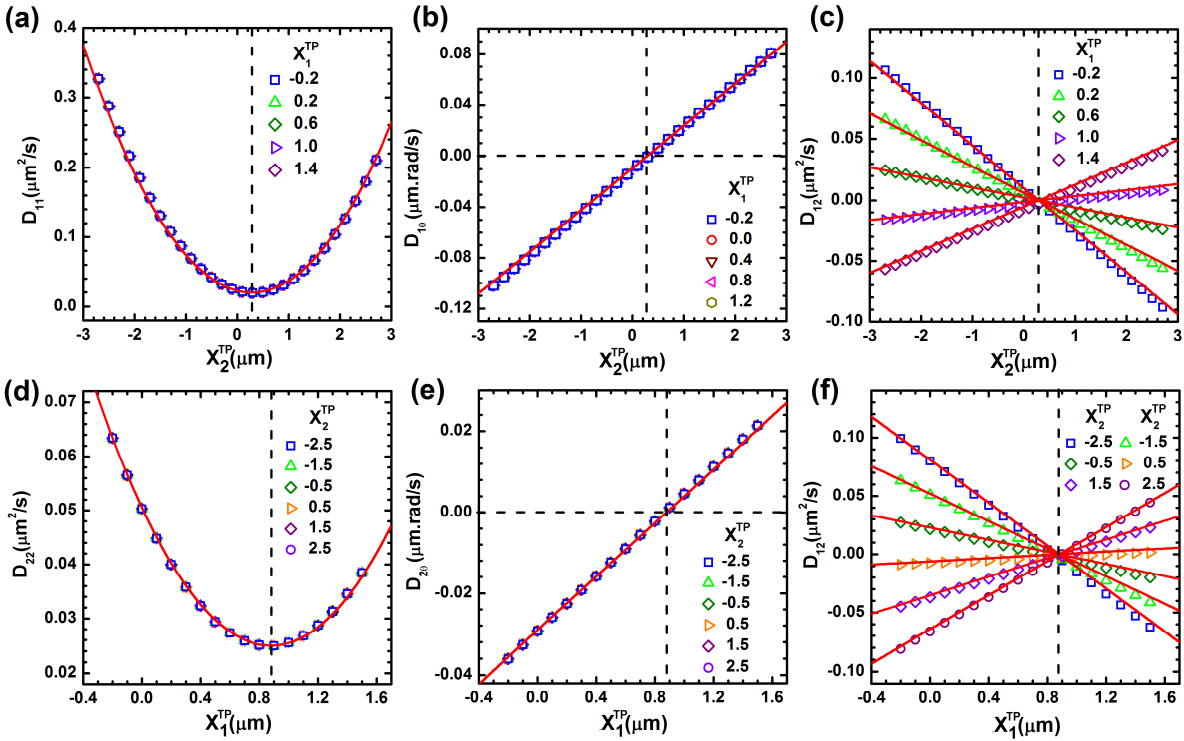

**Figure 4. Dependences of the diffusion coefficients on the TPs:** (a-f) variations of the measured diffusion coefficients ($D_{11}$, $D_{1\theta}$, $D_{12}$, $D_{22}$, $D_{2\theta}$, $D_{12}$) with the TP position $\left(X_1^{TP}, X_2^{TP}\right)$ with respect to the cross point of two arms. Different data symbols in (a-c) correspond to $X_1^{TP}$ fixed at different values (-0.2 μm, 0.2 μm, 0.6 μm, 1.0 μm and 1.4 μm); different data symbols in (d-f) correspond to $X_2^{TP}$ fixed at different values (-2.5 μm, -1.5 μm, -0.5 μm, 0.5 μm, 1.5 μm and 2.5 μm). Red solid lines are theoretical plots based on Eqs. 5a-5f by using the measured diffusion coefficients, the dashed lines indicate the location of the CoH at (0.88μm, 0.28μm) where $D_{11}$ and $D_{22}$ are at minima while $D_{1\theta}$, $D_{2\theta}$ and $D_{12}$ are zero.

Brenner classified particles as non-skewed or skewed based on whether their shape symmetry allows for the existence of CoH or not. Theoretical studies by Brenner and others show that in 3D only particles with three or more planes of symmetry are non-skewed, while particles with two or fewer planes of symmetry are skewed and their translational and rotational motions are intrinsically coupled. No previous work has discussed the 2D case. Our experimental results present the first experimental evidence that in 2D the CoH exists even for Brownian particles with no symmetry. In another word, particles are always non-skewed in 2D, independent of their shape. This is in sharp contrast to Brownian motion in 3D.

## ■ LANGEVIN THEORY

To understand the experimental results, we have developed an analytical model based on the Langevin theory. Our model is based on the fact that the asymmetric boomerangs confined in 2D possess a CoH as shown by the above experiments. We assume that the CoH is located at $(x_1^{CoH}, x_2^{CoH})$ and that the vector $\mathbf{r}$ from the CoH to the tracking point (TP) makes an angle $\varphi$ with the $X_1$ axis. We use $r$ to denote the distance between the TP and the CoH, so $\mathbf{r} = r\cos[\theta(t)+\varphi]\hat{x}_1 + r\sin[\theta(t)+\varphi]\hat{x}_2$. The TP position in the lab frame can be simply written as:

$$\begin{pmatrix} x_1(t) \\ x_2(t) \end{pmatrix} = \begin{pmatrix} x_1^{CoH}(t) \\ x_2^{CoH}(t) \end{pmatrix} + r\begin{pmatrix} \cos[\theta(t)+\varphi] \\ \sin[\theta(t)+\varphi] \end{pmatrix} \quad (1)$$

where $\theta(t)$ is the particle orientation. Since $\varphi$ is a constant, the velocity of the TP can be written as:

$$\begin{pmatrix} \dot{x}_1(t) \\ \dot{x}_2(t) \end{pmatrix} = \begin{pmatrix} \dot{x}_1^{CoH}(t) \\ \dot{x}_2^{CoH}(t) \end{pmatrix} + r\begin{pmatrix} -\sin[\theta(t)+\varphi] \\ \cos[\theta(t)+\varphi] \end{pmatrix} \dot{\theta}(t) \quad (2)$$

Since the displacements of the CoH and *r* are not correlated, $\langle [\Delta x_i^{CoH}(t)][\Delta r_{x_i}(t)] \rangle = 0$ and the mean displacements (MDs) and the mean square displacements (MSDs) of the TP can be expressed as:

$$\langle \Delta x_i(t) \rangle = \langle \Delta x_i^{CoH}(t) \rangle + \langle \Delta r_{x_i}(t) \rangle \tag{3}$$

$$\langle [\Delta x_i(t)]^2 \rangle = \langle [\Delta x_i^{CoH}(t)]^2 \rangle + \langle [\Delta r_{x_i}(t)]^2 \rangle \tag{4}$$

where $i = 1, 2$.

**Displacements of the CoH.** According to the definition of the CoH, the coupling diffusion coefficients and the coupling resistance coefficients are zero at CoH, or the diffusion and the resistance tensors for the CoH are diagonalized. Under no external force and over-damped conditions, the 2D Brownian motion of the CoH in the body frame can be described by the Langevin equation:

$$\begin{pmatrix} \zeta_{11}^{CoH} & 0 & 0 \\ 0 & \zeta_{22}^{CoH} & 0 \\ 0 & 0 & \zeta_{\theta\theta} \end{pmatrix} \begin{pmatrix} \dot{X}_1^{CoH}(t) \\ \dot{X}_2^{CoH}(t) \\ \dot{\theta}(t) \end{pmatrix} = \begin{pmatrix} \xi_1(t) \\ \xi_2(t) \\ \xi_\theta(t) \end{pmatrix} \tag{5}$$

where $\zeta_{ij}^{CoH}$ is the hydrodynamic resistance tensor. $\xi_i(t)$ is the Gaussian random force. According to the fluctuation-dissipation theorem, the resistance tensor is related to $\xi_i(t)$:

$$\langle \xi_i(t) \rangle = 0$$
$$\langle \xi_i(t) \xi_j(t') \rangle = 2k_B T \zeta_{ij}^{CoH} \delta(t-t')$$

where $i, j = 1, 2, \theta$. The diffusion tensor is related to the hydrodynamic resistance through the Einstein-Smoluchowski relationship: $D_{ij}^{CoH} = k_B T (\zeta_{ij}^{CoH})^{-1}$. Combining this with Eq. 5 then leads to $\dot{X}_i^{CoH} = \frac{1}{k_B T} D_{ij}^{CoH} \xi_j(t)$. To simplify the derivations, we scale the random noise by defining $\eta_i(t)$: $\xi_j(t) = k_B T \sqrt{2/D_{ij}^{CoH}} \eta_j(t)$, then $\langle \eta_i(t) \rangle = 0; \langle \eta_i(t)\eta_j(t') \rangle = \delta_{ij}\delta(t-t')$. Eq. 5 can then be rewritten as:

$$\begin{pmatrix} \dot{X}_1^{CoH}(t) \\ \dot{X}_2^{CoH}(t) \end{pmatrix} = \begin{pmatrix} \sqrt{2D_{11}^{CoH}} & 0 \\ 0 & \sqrt{2D_{22}^{CoH}} \end{pmatrix} \begin{pmatrix} \eta_1(t) \\ \eta_2(t) \end{pmatrix} \quad (6a)$$

$$\dot{\theta}(t) = \sqrt{2D_\theta}\eta_\theta(t) \quad (6b)$$

The velocities of the CoH in the lab can be obtained through a rotation transformation of the body frame velocities (Eq. 6a):

$$\begin{pmatrix} \dot{x}_1^{CoH}(t) \\ \dot{x}_2^{CoH}(t) \end{pmatrix} = \begin{pmatrix} \cos\theta(t) & -\sin\theta(t) \\ \sin\theta(t) & \cos\theta(t) \end{pmatrix} \begin{pmatrix} \sqrt{2D_{11}^{CoH}} & 0 \\ 0 & \sqrt{2D_{22}^{CoH}} \end{pmatrix} \begin{pmatrix} \eta_1(t) \\ \eta_2(t) \end{pmatrix} \quad (7)$$

Eqs. 6-7 are essentially the same as the Langevin equations for an ellipsoidal particle with its center of mass as the TP. The solutions can be obtained similarly to Ref. [47]. So the MDs of the CoH for fixed initial angle $\theta_0$ go to zero:

$$\langle \Delta x_1^{CoH}(t) \rangle_{\theta_0} = \langle \Delta x_2^{CoH}(t) \rangle_{\theta_0} = 0 \quad (8)$$

The MSDs of the CoH with fixed initial orientation $\theta_0$ can be expressed as [47]:

$$\langle [\Delta x_1^{CoH}(t)]^2 \rangle_{\theta_0} = 2\overline{D}^{CoH} t - \cos 2\theta_0 \frac{\Delta D^{CoH}}{4D_\theta}\left(1 - e^{-4D_\theta t}\right) \quad (9a)$$

$$\langle [\Delta x_2^{CoH}(t)]^2 \rangle_{\theta_0} = 2\overline{D}^{CoH} t + \cos 2\theta_0 \frac{\Delta D^{CoH}}{4D_\theta}\left(1 - e^{-4D_\theta t}\right) \quad (9b)$$

where $\overline{D}^{CoH} = (D_{11}^{CoH} + D_{22}^{CoH})/2$, $\Delta D = D_{22}^{CoH} - D_{11}^{CoH}$.

**Displacements of the Vector r.** The displacements of *r* along the $x_1$ and $x_2$ directions for small angular displacement $\Delta\theta$ can be expressed as:

$$\Delta r_{x_1}(t) = r\cos[\theta_0 + \varphi + \Delta\theta(t)] - r\cos(\theta_0 + \varphi) \tag{10a}$$

$$\Delta r_{x_2}(t) = r\sin[\theta_0 + \varphi + \Delta\theta(t)] - r\sin(\theta_0 + \varphi) \tag{10b}$$

The random variations of $\Delta\theta(t)$ have a Gaussian distribution with a zero mean. It can be shown that for an integer $n$, $\langle \sin[n\Delta\theta(t)]\rangle = 0$, and $\langle \cos[n\Delta\theta(t)]\rangle = \exp(-n^2 D_\theta t)$. Then using Eq. 10a-b, the MDs of the vector *r* can be calculated as:

$$\langle \Delta r_{x_1}(t)\rangle_{\theta_0} = -r\cos(\theta_0 + \varphi)\left(1 - e^{-D_\theta t}\right) \tag{11a}$$

$$\langle \Delta r_{x_2}(t)\rangle_{\theta_0} = -r\sin(\theta_0 + \varphi)\left(1 - e^{-D_\theta t}\right) \tag{11b}$$

and the MSDs of *r* can be calculated as:

$$\langle [\Delta r_{x_1}(t)]^2\rangle_{\theta_0} = 2r^2 \cos^2(\theta_0 + \varphi)\left(1 - e^{-D_\theta t}\right) - \frac{r^2}{2}\cos 2(\theta_0 + \varphi)\left(1 - e^{-4D_\theta t}\right) \tag{12a}$$

and:

$$\langle [\Delta r_{x_2}(t)]^2\rangle_{\theta_0} = 2r^2 \sin^2(\theta_0 + \varphi)\left(1 - e^{-D_\theta t}\right) + \frac{1}{2}r^2 \cos 2(\theta_0 + \varphi)\left(1 - e^{-4D_\theta t}\right) \tag{12b}$$

**MDs and MSDs of the TP in the Lab Frame.** The MDs of the TP for fixed initial angle $\theta_0$ can be obtained by substituting Eq. 8 and Eq. 11 (a-b) into Eq. 3:

$$\langle \Delta x_1(t)\rangle_{\theta_0} = -r\cos(\theta_0 + \varphi)\left(1 - e^{-D_\theta t}\right) \tag{13a}$$

$$\langle \Delta x_2(t)\rangle_{\theta_0} = -r\sin(\theta_0 + \varphi)\left(1 - e^{-D_\theta t}\right) \tag{13b}$$

The MSDs of the TP for the initial orientation fixed at $\theta_0$ can be obtained similarly by substituting Eq. 9 (a-b) and Eq. 12 (a-b) into Eq. 4:

$$\left\langle [\Delta x_1(t)]^2 \right\rangle_{\theta_0} = 2\overline{D}^{CoH}t + 2r^2 \cos^2(\theta_0 + \varphi)\left(1 - e^{-D_\theta t}\right)$$
$$- \left(\frac{\Delta D}{4D_\theta}\cos 2\theta_0 + \frac{r^2}{2}\cos 2(\theta_0 + \varphi)\right)\left(1 - e^{-4D_\theta t}\right) \quad (14a)$$

$$\left\langle [\Delta x_2(t)]^2 \right\rangle_{\theta_0} = 2\overline{D}^{CoH}t + 2r^2 \sin^2(\theta_0 + \varphi)\left(1 - e^{-D_\theta t}\right)$$
$$+ \left(\frac{\Delta D}{4D_\theta}\cos 2\theta_0 + \frac{r^2}{2}\cos 2(\theta_0 + \varphi)\right)\left(1 - e^{-4D_\theta t}\right) \quad (14b)$$

For $\theta_0 = 0$, the MDs and the MSDs can be simplified from Eq. 13(a-b) as:

$$\langle \Delta x_1(t) \rangle_{\theta_0 = 0} = -r_1\left(1 - e^{-D_\theta t}\right) \quad (15a)$$

$$\langle \Delta x_2(t) \rangle_{\theta_0 = 0} = -r_2\left(1 - e^{-D_\theta t}\right) \quad (15b)$$

$$\left\langle [\Delta x_1(t)]^2 \right\rangle_{\theta_0 = 0} = 2\overline{D}^{CoH}t + 2r^2 \cos^2\varphi\left(1 - e^{-D_\theta t}\right) - \left(\frac{\Delta D}{4D_\theta} + \frac{r^2}{2}\cos(2\varphi)\right)\left(1 - e^{-4D_\theta t}\right) \quad (16a)$$

$$\left\langle [\Delta x_2(t)]^2 \right\rangle_{\theta_0 = 0} = 2\overline{D}^{CoH}t + 2r^2 \sin^2\varphi\left(1 - e^{-D_\theta t}\right) + \left(\frac{\Delta D}{4D_\theta} + \frac{r^2}{2}\cos(2\varphi)\right)\left(1 - e^{-4D_\theta t}\right) \quad (16b)$$

where $r_1 = r\cos\varphi, r_2 = r\sin\varphi$. The angle $\varphi$ between $\mathbf{r}$ and the $X_1$ axis is given by $\varphi = \tan^{-1}(r_2/r_1)$.

By averaging Eq. 15a-b and 16a-b over all possible initial orientations $\theta_0$, we obtain the angle averaged MDs and MSDs as:

$$\langle \Delta x_1(t) \rangle = \langle \Delta x_2(t) \rangle = 0 \quad (17)$$

$$\left\langle [\Delta x_1(t)]^2 \right\rangle = \left\langle [\Delta x_2(t)]^2 \right\rangle = 2\overline{D}^{CoH}t + r^2\left(1 - e^{-D_\theta t}\right) \quad (18)$$

Therefore, the lab-frame short-time diffusion coefficient is $\overline{D}^{ST} = \overline{D}^{CoH} + r^2 D_\theta/2$ and the lab frame long-time diffusion coefficient is $\overline{D}^{LT} = \overline{D}^{CoH}$.

**MDs, MSDs and Displacement Correlations in the Body Frame.** To calculate the MDs, MSDs and displacement correlations in the body frame, we follow the experimental procedure and start with the lab frame. The velocities in the body frame are given by the rotation transformation of the lab frame velocities:

$$\dot{X}_1(t) = \cos\theta'(t)\dot{x}_1(t) + \sin\theta'(t)\dot{x}_2(t)$$
$$\dot{X}_2(t) = -\sin\theta'(t)\dot{x}_1(t) + \cos\theta'(t)\dot{x}_2(t) \quad (19)$$

It should be noted that we have used here $\theta'(t)$ as the angle for the frame transformation as in experiments it is different from the instantaneous particle orientation $\theta(t)$. Using Eq. 2 and Eq. 6b, we have:

$$\dot{X}_1(t) = \dot{X}_1^{CoH}(t) - \sin[\theta(t)+\varphi-\theta'(t)]r\alpha_\theta\eta_\theta(t) = \dot{X}_1^{CoH}(t) + \dot{R}_1(t)$$
$$\dot{X}_2(t) = \dot{X}_2^{CoH}(t) + \cos[\theta(t)+\varphi-\theta'(t)]r\alpha_\theta\eta_\theta(t) = \dot{X}_2^{CoH}(t) + \dot{R}_2(t) \quad (20)$$

where $\dot{R}_1(t)$ and $\dot{R}_2(t)$ are the velocities of the vector *r* transformed into the body frame. Considering that the motion of the CoH and that of $R_1$ or $R_2$ are not correlated, we can separate the MDs, MSDs and displacement correlations in the body frame into two terms:

$$\langle \Delta X_i(t) \rangle = \langle \Delta X_i^{CoH}(t) \rangle + \langle \Delta R_i(t) \rangle \quad (21)$$

$$\langle [\Delta X_i(t)]^2 \rangle = \langle [\Delta X_i^{CoH}(t)]^2 \rangle + \langle [\Delta R_i(t)]^2 \rangle \quad (22)$$

$$\langle \Delta X_i(t)\Delta\theta(t) \rangle = \langle \Delta X_i^{CoH}(t)\Delta\theta(t) \rangle + \langle \Delta R_i(t)\Delta\theta(t) \rangle \quad (23)$$

$$\langle \Delta X_i(t)\Delta X_j(t) \rangle = \langle \Delta R_i(t)\Delta R_j(t) \rangle \quad (24)$$

From the definition we have:

$$\dot{R}_1(t) = -\sin[\theta(t)+\varphi+\theta'(t)]r\alpha_\theta\eta_\theta(t) \quad (25a)$$

$$\dot{R}_2(t) = \cos[\theta(t)+\varphi-\theta'(t)]r\alpha_\theta\eta_\theta(t) \quad (25b)$$

For a single time step between $t_0$ and $t_0+\tau$, the displacements of $R_1$ and $R_2$ are respectively

$$\Delta R_1(\tau) = -\int_{t_0}^{t_0+\tau} dt' \sin[\theta(t')+\varphi-\theta']r\alpha_\theta\eta_\theta(t') \text{ and } \Delta R_2(\tau) = \int_{t_0}^{t_0+\tau} dt' \cos[\theta(t')+\varphi-\theta']r\alpha_\theta\eta_\theta(t').$$ For

small $\tau$, $\theta(t')-\theta'$ is small, hence we can use the Taylor expansion:

$$\Delta R_1(\tau) = -\sin\varphi \int_{t_0}^{t_0+\tau} dt' r\alpha_\theta\eta_\theta(t') - \cos\varphi \int_{t_0}^{t_0+\tau} dt'[\theta(t')-\theta']r\alpha_\theta\eta_\theta(t') \tag{26a}$$

$$\Delta R_2(\tau) = \cos\varphi \int_{t_0}^{t_0+\tau} dt' r\alpha_\theta\eta_\theta(t') - \sin\varphi \int_{t_0}^{t_0+\tau} dt'[\theta(t')-\theta']r\alpha_\theta\eta_\theta(t') \tag{26b}$$

In the continuous body frame (CBF), $\theta' = [\theta(t_0)+\theta(t_0+\tau)]/2$.[54] Using the expressions

$$\theta(t_0) = \theta(t') - \int_{t_0}^{t'} dt'' \dot\theta(t'') \text{ and } \theta(t_0+\tau) = \theta(t') + \int_{t'}^{t_0+\tau} dt'' \dot\theta(t''), \text{ we have}$$

$$\theta' = \theta(t') + \frac{\alpha_\theta}{2}\left[\int_{t'}^{t_0+\tau} dt'' \eta_\theta(t'') - \int_{t_0}^{t'} dt'' \eta_\theta(t'')\right].$$ Using this expression of $\theta'$ into Eq. 26a and 26b, it

can be shown that the second terms in Eq. 26a-b is zero, i.e. $\int_{t_0}^{t_0+\tau} dt'[\theta(t')-\theta'(t')]r\alpha_\theta\eta_\theta(t') = 0$.

Then the MDs for $R_1$ and $R_2$ can be calculated as:

$$\langle \Delta R_1(\tau) \rangle = 0 \tag{27a}$$

$$\langle \Delta R_2(\tau) \rangle = 0 \tag{27b}$$

And the MSDs can be obtained as:

$$\langle [\Delta R_1(\tau)]^2 \rangle = 2r^2 \sin^2\varphi D_\theta \tau = 2r_2^2 D_\theta \tau \tag{28a}$$

$$\langle [\Delta R_2(\tau)]^2 \rangle = 2r^2 \cos^2\varphi D_\theta \tau = 2r_1^2 D_\theta \tau \tag{28b}$$

The coupling of displacements of $R_1$ and $R_2$ with the rotational displacements can be expressed as:

$$\langle \Delta R_1(\tau)\Delta\theta(\tau)\rangle = -2r\sin\varphi D_\theta \tau = -2r_2 D_\theta \tau \tag{29a}$$

$$\langle \Delta R_2(\tau)\Delta\theta(\tau)\rangle = 2r\cos\varphi D_\theta \tau = 2r_1 D_\theta \tau \tag{29b}$$

$$\langle \Delta R_1(\tau)\Delta R_2(\tau)\rangle = -2r^2 \sin\varphi\cos\varphi D_\theta \tau = -2r_1 r_2 D_\theta \tau \tag{29c}$$

For $t = n\tau$ with $n$ being an integer, the displacements in the body frame are accumulations of the displacements in individual time intervals: $\Delta R_{1,2}(t,t_0) = \sum_{i=0}^{n-1}\Delta R_{1,2}(\tau,t_0+i\tau)$. The MDs, the MSDs and the displacement correlations are then expressed as:

$$\langle \Delta R_{1,2}(t)\rangle = \left\langle \sum_{i=0}^{n-1}\Delta R_{1,2}(\tau,t_0+i\tau)\right\rangle = n\langle \Delta R_{1,2}(\tau)\rangle \tag{30a}$$

$$\langle [\Delta R_{1,2}(t)]^2\rangle = \left\langle \sum_{i=0}^{n-1}[\Delta R_{1,2}(\tau,t_0+i\tau)]^2\right\rangle + \left\langle \sum_{i\neq j}[\Delta R_{1,2}(\tau,t_0+i\tau)][\Delta R_{1,2}(\tau,t_0+j\tau)]\right\rangle$$
$$= n\langle [\Delta R_{1,2}(\tau)]^2\rangle + (n^2-n)\langle [\Delta R_{1,2}(\tau)]\rangle^2 \tag{30b}$$

$$\langle \Delta R_{1,2}(t)\Delta\theta(t)\rangle = \left\langle \left[\sum_{i=0}^{n-1}\Delta R_{1,2}(\tau,t_0+i\tau)\right]\left[\sum_{j=0}^{n-1}\Delta\theta(\tau,t_0+j\tau)\right]\right\rangle = n\langle \Delta R_{1,2}(\tau)\Delta\theta(\tau)\rangle \tag{30c}$$

$$\langle \Delta R_1(t)\Delta R_2(t)\rangle = \left\langle \sum_{i=0}^{n-1}\Delta R_1(\tau,t_0+i\tau)\Delta R_2(\tau,t_0+i\tau)\right\rangle + \left\langle \sum_{i\neq j}[\Delta R_1(\tau,t_0+i\tau)][\Delta R_2(\tau,t_0+j\tau)]\right\rangle$$
$$= n\langle \Delta R_1(\tau)\Delta R_2(\tau)\rangle + (n^2-n)\langle [\Delta R_1(\tau)]\rangle\langle [\Delta R_2(\tau)]\rangle \tag{30d}$$

By combining Eq. 30(a-d) with Eq. 28(a-b) and 29(a-b) we get:

$$\langle [\Delta R_1(\tau)]^2\rangle = 2r_2^2 D_\theta t \tag{31a}$$

$$\langle [\Delta R_2(\tau)]^2\rangle = 2r_1^2 D_\theta t \tag{31b}$$

$$\langle \Delta R_1(\tau)\Delta\theta(\tau)\rangle = -2r_2 D_\theta t \tag{31c}$$

$$\langle \Delta R_2(\tau) \Delta\theta(\tau)\rangle = 2r_1 D_\theta t \tag{31d}$$

$$\langle \Delta R_1(\tau) \Delta R_2(\tau)\rangle = -2r_1 r_2 D_\theta t \tag{31e}$$

By substituting 27(a-b) and 31(a-e) into Eqs. 21-24, we get the MDs, the MSDs and the cross correlations in the body frame as:

$$\langle \Delta X_1(t)\rangle = 0 \tag{32a}$$

$$\langle \Delta X_2(t)\rangle = 0 \tag{32b}$$

$$\langle [X_1(t)]^2\rangle = 2\left(D_{11}^{CoH} + r_2^2 D_\theta\right) t \tag{32c}$$

$$\langle [X_2(t)]^2\rangle = 2\left(D_{22}^{CoH} + r_1^2 D_\theta\right) t \tag{32d}$$

$$\langle X_1(t)\Delta\theta(t)\rangle = -2r_2 D_\theta t \tag{32e}$$

$$\langle X_2(t)\Delta\theta(t)\rangle = 2r_1 D_\theta t \tag{32f}$$

$$\langle X_1(t) X_2(t)\rangle = -2r_1 r_2 D_\theta t \tag{32g}$$

So the anisotropic diffusion coefficients are:

$$D_{11} = D_{11}^{CoH} + r_2^2 D_\theta \tag{33a}$$

$$D_{22} = D_{22}^{CoH} + r_1^2 D_\theta \tag{33b}$$

$$D_{1\theta} = -r_2 D_\theta \tag{33c}$$

$$D_{2\theta} = r_1 D_\theta \tag{33d}$$

$$D_{12} = -r_1 r_2 D_\theta \tag{33e}$$

Therefore from the measured diffusion coefficients with any TP, we can find the location of the CoH using $r_1 = D_{2\theta}/D_\theta$, $r_2 = -D_{1\theta}/D_\theta$, $r = \sqrt{D_{1\theta}^2 + D_{2\theta}^2}/D_\theta$ and $\varphi = \tan^{-1}(-D_{1\theta}/D_{2\theta})$. The formulae for symmetric boomerangs can be recovered by setting $\varphi = \pi$.

■ **DISCUSSIONS**

With respect to the origin of the body frame as defined in the experiments, we denote the positions of the CoH and an arbitrary TP respectively as $(d_1, d_2)$ and $(X_1^{TP}, X_2^{TP})$, then we have $r_1 = X_1^{TP} - d_1$, $r_2 = X_2^{TP} - d_2$. With these notations, we compare the theoretical model firstly with the experimental results where the cross point of the two arms is used as the TP. Based on the measured values of $D_\theta$, $D_{1\theta}$, and $D_{2\theta}$, the distance between the TP and the CoH can be determined through Eq. 33c-d: $d_1 = 0.88$ μm, $d_2 = 0.28$ μm. With these values, Eqs. 15a-b can be plotted with no fitting parameters, which show good agreements with the experimental data (Figure 2d). The MSDs in the lab frame can be well fitted with Eqs. 16a-b and 18. By setting $D_\theta$ as 0.034 rad²/s, best fitting gives $D^{CoH} = 0.0225\ \mu m^2/s$ and $r = 0.92$ μm (Fig. 2a and 2c).

We then compare the theoretical expressions with the experiments regarding the dependences of the diffusion coefficients on the TPs. We plot $D_{11} = D_{11}^{CoH} + (X_2^{TP} - d_2)^2 D_\theta$, $D_{1\theta} = (X_2^{TP} - d_2) D_\theta$, and $D_{12} = (X_1^{TP} - d_1)(X_2^{TP} - d_2) D_\theta$ using the measured values of $d_1$, $d_2$ and $D_\theta$ and only $D_{11}^{CoH}$ as the fitting parameter. The agreements between theory and experiments are excellent. Similar agreements are also obtained for the $X_1$ dependence of $D_{22}$ and $D_{2\theta}$. The best fittings yield $D_{11}^{CoH} = 0.02\ \mu m^2/s$ and $D_{22}^{CoH} = 0.025\ \mu m^2/s$ which are consistent with the short and long time diffusion coefficients $D^{ST} = 0.0375$ μm²/s and $D^{LT} = 0.0225$ μm²/s obtained from the lab frame MSDs.

The symmetry of 2D geometric shapes can be characterized by two families of discrete point groups: the cyclic group ($C_1$, $C_2$, …$C_n$), and the dihedral group ($D_1$, $D_2$, …$D_n$). For particles with rotational symmetry $C_n$ (n>1), their CoM overlaps with their rotation symmetry center. When CoM is chosen as the TP, then the diffusion coefficient tensor should remain unchanged when the body frame is rotated with respect to the particle by an angle of $\theta_n = 2\pi/n$, or

$$R(\theta_n) \begin{pmatrix} D_{11} & D_{12} & D_{1\theta} \\ D_{12} & D_{22} & D_{2\theta} \\ D_{1\theta} & D_{2\theta} & D_{\theta} \end{pmatrix} R^{-1}(\theta_n) = \begin{pmatrix} D_{11} & D_{12} & D_{1\theta} \\ D_{12} & D_{22} & D_{2\theta} \\ D_{1\theta} & D_{2\theta} & D_{\theta} \end{pmatrix}$$

where $R(\theta_n) = \begin{pmatrix} \cos\theta_n & \sin\theta_n & 0 \\ -\sin\theta_n & \cos\theta_n & 0 \\ 0 & 0 & 1 \end{pmatrix}$ is the rotation transformation matrix. For particles with $C_2$ symmetry, the above requirement necessitates that only the diagonal elements of the diffusion tensor is non-zero, or the diffusion behavior is similar to that of an ellipsoids. For particles with $C_n$ (n > 2) symmetry, it can be easily seen that when the CoM is conveniently used as the TP, their diffusion behaviors are similar to those of circular disks described by one translation and one rotation coefficients. Given that the $C_n$ symmetry is contained in the $D_n$ symmetry, particles of $D_n$ symmetry with n > 2 behave similarly as circular disks. With the above results for non-symmetry boomerangs, we can categorize the dependence of 2D Brownian motion on the particle shapes into 4 different cases with diffusion behaviors similar respectively to circular disks, elliptical disks, symmetric boomerangs and asymmetric boomerangs (Table 1).

| Particle Symmetry | $D_\infty$ | $D_3…D_n$ | $C_3…C_n$ | $D_2$, $C_2$ | $D_1$ | $C_1$ |
|---|---|---|---|---|---|---|
| Exemplary shapes | Circular disk | Regular Polygons | Swastika ($C_4$) | Elliptic disk ($D_2$), dollar sign ($C_2$) | Symmetric boomerang | Asymmetric boomerang |
| Normal TP | CoM | | | CoM | On symmetry line | Any point |

| Location of CoH | CoH=CoM | CoH=CoM | CoH≠CoM | CoH≠CoM |
|---|---|---|---|---|
| Non-zero Diffusion coefficients | $D, D_\theta$ | $D_{11}, D_{22}, D_\theta$ | $D_{11}, D_{22}, D_{1\theta}, D_\theta$ | $D_{11}, D_{22}, D_{12}, D_{1\theta}, D_{2\theta}, D_\theta$ |
| Independent diffusion coefficients | $D, D_\theta$ | $D_{11}, D_{22}, D_\theta$ | $D_{11}, D_{22}, D_{1\theta}, D_\theta$ | $D_{11}, D_{22}, D_{1\theta}, D_{2\theta}, D_\theta$ |
| MSDs | linear with $t$ for all time | | linear with $t$ only at short and long time | |
| Crossover time | None | | $\sim 1/2D_\theta$ | |
| Displacement PDF | Gaussian | | Non-Gaussian | |
| References | Ref. [4] | | Ref [47] | Ref [54] | This paper |

## ■CONCLUSION

To summarize, we have shown that for single non-symmetric colloidal particles confined in quasi-two dimensions, the MSDs grow linearly with time only at short and long time with different diffusion coefficients and the MDs for fixed initial angle are biased and non-zero because of translation-rotation coupling. The diffusion coefficients measured through single particle motion tracking depend strongly on the position of TPs. In particular, for 2D Brownian motion of non-symmetric particles, there always exists a unique tracking point, i.e. the CoH, where translation and rotation can be decoupled and the MSDs and MDs behave similarly as those of elliptical disks. For that reason, among the 6 diffusion coefficients only 5 are independent. Based on these results, we are able to categorize the behavior of 2D Brownian motion of arbitrarily shaped particles based on their shape symmetry.

Single particle tracking in both 2D and 3D is not only widely used in studying colloid systems[11-12, 44-45] but also a powerful tool in studying the dynamics of molecules in biological systems such as membrane dynamics[56] and intra/intercellular molecule transport.[57-59] Fluorescent tagging is commonly used for tracking macromolecules and cells. Because these fluorescent tags are not necessarily coincident with the CoH, our results indicate that the translation-rotation coupling has significant impact on the behavior of measured MSDs and on the measurements of particle diffusivities. Although particles of arbitrary shapes in 2D are shown to be non-skewed, the effects of the translation-rotation coupling are still important because the CoH for non-symmetric particles is unknown before the diffusion behaviors of any TP are analyzed in experiments.

**Corresponding Author**

\* Corresponding authors: qwei@kent.edu; jselinge@kent.edu.

**Author Contributions:**

†: These two authors contribute equally.

**Acknowledgements**

The work was supported by a Farris Family Award and partially supported by NSF ECCS-0824175 and NSF DMR-1106014.

# REFERENCES

1. Chandrasekhar, S., Stochastic Problems in Physics and Astronomy. *Reviews of Modern Physics* **1943,** *15* (1), 1-89.

2. Hanggi, P.; Marchesoni, F., Introduction: 100 Years of Brownian Motion. *Chaos* **2005,** *15* (2), 026101.

3. Frey, E.; Kroy, K., Brownian Motion: A Paradigm of Soft Matter and Biological Physics. *Annalen Der Physik* **2005,** *14* (1-3), 20-50.

4. Einstein, A., On the Movement of Small Particles Suspended in Stationary Liquids Required by the Molecular-Kinetic Theory of Heat. *Annals of Physics* **1905,** *17*, 549-560.

5. Franosch, T.; Grimm, M.; Belushkin, M.; Mor, F. M.; Foffi, G.; Forro, L.; Jeney, S., Resonances Arising from Hydrodynamic Memory in Brownian Motion. *Nature* **2011,** *478* (7367), 85-88.

6. Li, T. C.; Kheifets, S.; Medellin, D.; Raizen, M. G., Measurement of the Instantaneous Velocity of a Brownian Particle. *Science* **2010,** *328* (5986), 1673-1675.

7. Huang, R. X.; Chavez, I.; Taute, K. M.; Lukic, B.; Jeney, S.; Raizen, M. G.; Florin, E. L., Direct Observation of the Full Transition from Ballistic to Diffusive Brownian Motion in a Liquid. *Nature Physics* **2011,** *7* (7), 576-580.

8. Romanczuk, P.; Schimansky-Geier, L., Brownian Motion with Active Fluctuations. *Physical Review Letters* **2011,** *106* (23), 230601-230601.

9. Turiv, T.; Lazo, I.; Brodin, A.; Lev, B. I.; Reiffenrath, V.; Nazarenko, V. G.; Lavrentovich, O. D., Effect of Collective Molecular Reorientations on Brownian Motion of Colloids in Nematic Liquid Crystal. *Science* **2013,** *342* (6164), 1351-1354.

10. Mason, T. G.; Weitz, D. A., Optical Measurements of Frequency-Dependent Linear Viscoelastic Moduli of Complex Fluids. *Physical Review Letters* **1995,** *74* (7), 1250-1253.

11. Mukhopadhyay, A.; Granick, S., Micro- and Nanorheology. *Current Opinion in Colloid & Interface Science* **2001,** *6* (5-6), 423-429.

12. Waigh, T. A., Microrheology of Complex Fluids. *Reports on Progress in Physics* **2005,** *68* (3), 685-742.

13. Chou, C. F.; Bakajin, O.; Turner, S. W. P.; Duke, T. A. J.; Chan, S. S.; Cox, E. C.; Craighead, H. G.; Austin, R. H., Sorting by Diffusion: An Asymmetric Obstacle Course for Continuous Molecular Separation. *Proc. Natl. Acad. Sci. U. S. A.* **1999,** *96* (24), 13762-13765.

14. Romanczuk, P.; Baer, M.; Ebeling, W.; Lindner, B.; Schimansky-Geier, L., Active Brownian Particles from Individual to Collective Stochastic Dynamics. *European Physical Journal-Special Topics* **2012,** *202* (1), 1-162.

15. Zerrouki, D.; Rotenberg, B.; Abramson, S.; Baudry, J.; Goubault, C.; Leal-Calderon, F.; Pine, D. J.; Bibette, M., Preparation of Doublet, Triangular, and Tetrahedral Colloidal Clusters by Controlled Emulsification. *Langmuir* **2006,** *22* (1), 57-62.

16. Glotzer, S. C.; Solomon, M. J., Anisotropy of Building Blocks and Their Assembly into Complex Structures. *Nature Materials* **2007,** *6* (8), 557-562.

17. Sacanna, S.; Pine, D. J., Shape-Anisotropic Colloids: Building Blocks for Complex Assemblies. *Current Opinion in Colloid & Interface Science* **2011,** *16* (2), 96-105.

18. Solomon, M. J., Directions for Targeted Self-Assembly of Anisotropic Colloids from Statistical Thermodynamics. *Current Opinion in Colloid & Interface Science* **2011,** *16* (2), 158-167.

19. Damasceno, P. F.; Engel, M.; Glotzer, S. C., Predictive Self-Assembly of Polyhedra into Complex Structures. *Science* **2012,** *337* (6093), 453-457.

20. Sacanna, S.; Irvine, W. T. M.; Chaikin, P. M.; Pine, D. J., Lock and Key Colloids. *Nature* **2010,** *464* (7288), 575-578.

21. Chen, Q.; Bae, S. C.; Granick, S., Directed Self-Assembly of a Colloidal Kagome Lattice. *Nature* **2011,** *469* (7330), 381-384.

22. Wang, J.; Byrne, J. D.; Napier, M. E.; DeSimone, J. M., More Effective Nanomedicines through Particle Design. *Small* **2011,** *7* (14), 1919-1931.

23. Brenner, H., Coupling between the Translational and Rotational Brownian Motions of Rigid Particles of Arbitrary Shape. *Journal of Colloid Science* **1955,** *20*, 104-122.

24. Brenner, H., The Stokes Resistance of an Arbitrary Particle-Ii an Extension. *Chemical Engineering Science* **1964,** *19*, 599-629.

25. Brenner, H., Coupling between the Translational and Rotational Brownian Motion of Rigid Particles of Arbitrary Shape. *Journal Of Colloid and Interface Science* **1967,** *23*, 407-436.

26. Garcia De La Torre, J.; Bloomfield, V. A., Hydrodynamic Properties of Macromolecular Complexes .1. Translation. *Biopolymers* **1977,** *16* (8), 1747-1763.

27. Garcia De La Torre, J.; Bloomfield, V. A., Hydrodynamics of Macromolecular Complexes .2. Rotation. *Biopolymers* **1977,** *16* (8), 1765-1778.

28. Wegener, W. A., Diffusion-Coefficients for Rigid Macromolecules with Irregular Shapes That Allow Rotational-Translational Coupling. *Biopolymers* **1981,** *20* (2), 303-326.

29. Dickinson, E.; Allison, S. A.; McCammon, J. A., Brownian Dynamics with Rotation Translation Coupling. *Journal of the Chemical Society-Faraday Transactions Ii* **1985,** *81* (APR), 591-601.

30. Squires, T. M.; Bazant, M. Z., Breaking Symmetries in Induced-Charge Electro-Osmosis and Electrophoresis. *Journal of Fluid Mechanics* **2006,** *560*, 65.

31. Makino, M.; Doi, M., Sedimentation of a Particle with Translation-Rotation Coupling. *Journal of the Physical Society of Japan* **2003,** *72* (11), 2699-2701.

32. Doi, M.; Makino, M., Sedimentation of Particles of General Shape. *Physics of Fluids* **2005,** *17* (4), 043601.

33. Grima, R.; Yaliraki, S. N., Brownian Motion of an Asymmetrical Particle in a Potential Field. *Journal of Chemical Physics* **2007,** *127* (8), 084511.

34. Wittkowski, R.; Loewen, H., Self-Propelled Brownian Spinning Top: Dynamics of a Biaxial Swimmer at Low Reynolds Numbers. *Phys. Rev. E* **2012,** *85* (2), 021406.

35. Gibbs, J. G.; Kothari, S.; Saintillan, D.; Zhao, Y. P., Geometrically Designing the Kinematic Behavior of Catalytic Nanomotors. *Nano Letters* **2011,** *11* (6), 2543-2550.

36. Kummel, F.; ten Hagen, B.; Wittkowski, R.; Buttinoni, I.; Eichhorn, R.; Volpe, G.; Lowen, H.; Bechinger, C., Circular Motion of Asymmetric Self-Propelling Particles. *Physical Review Letters* **2013,** *110* (19).

37. Wensink, H. H.; Loewen, H.; Marechal, M.; Hartel, A.; Wittkowski, R.; Zimmermann, U.; Kaiser, A.; Menzel, A. M., Differently Shaped Hard Body Colloids in Confinement: From Passive to Active Particles. *European Physical Journal-Special Topics* **2013,** *222* (11), 3023-3037.

38. Mijalkov, M.; Volpe, G., Sorting of Chiral Microswimmers. *Soft Matter* **2013,** *9* (28), 6376-6381.

39. Schamel, D.; Pfeifer, M.; Gibbs, J. G.; Miksch, B.; Mark, A. G.; Fischer, P., Chiral Colloidal Molecules and Observation of the Propeller Effect. *Journal of the American Chemical Society* **2013,** *135* (33), 12353-12359.

40. Aristov, M.; Eichhorn, R.; Bechinger, C., Separation of Chiral Colloidal Particles in a Helical Flow Field. *Soft Matter* **2013,** *9* (8), 2525-2530.

41. Buttinoni, I.; Bialke, J.; Kuemmel, F.; Loewen, H.; Bechinger, C.; Speck, T., Dynamical Clustering and Phase Separation in Suspensions of Self-Propelled Colloidal Particles. *Physical Review Letters* **2013,** *110* (23), 238301.

42. Nguyen, N. H. P.; Klotsa, D.; Engel, M.; Glotzer, S. C., Emergent Collective Phenomena in a Mixture of Hard Shapes through Active Rotation. *Physical Review Letters* **2014,** *112* (7), 075701.

43. Wensink, H. H.; Kantsler, V.; Goldstein, R. E.; Dunkel, J., Controlling Active Self-Assembly through Broken Particle-Shape Symmetry. *Phys. Rev. E* **2014,** *89* (1), 010302.

44. Crocker, J. C.; Grier, D. G., Methods of Digital Video Microscopy for Colloidal Studies. *J Colloid Interface Sci,* **1996,** *179* 298-310.

45. Prasad, V.; Semwogerere, D.; Weeks, E. R., Confocal Microscopy of Colloids. *Journal of Physics-Condensed Matter* **2007,** *19* (11).

46. Maeda, H.; Maeda, Y., Direct Observation of Brownian Dynamics of Hard Colloidal Nanorods. *Nano Letters* **2007,** *7* (11), 3329-3335.

47. Han, Y.; Alsayed, A. M.; Nobili, M.; Zhang, J.; Lubensky, T. C.; Yodh, A. G., Brownian Motion of an Ellipsoid. *Science* **2006,** *314* (5799), 626-630.
28

48. Mukhija, D.; Solomon, M. J., Translational and Rotational Dynamics of Colloidal Rods by Direct Visualization with Confocal Microscopy. *Journal Of Colloid and Interface Science* **2007,** *314* (1), 98-106.

49. Anthony, S. M.; Kim, M.; Granick, S., Translation-Rotation Decoupling of Colloidal Clusters of Various Symmetries. *Journal of Chemical Physics* **2008,** *129* (24), 244701.

50. Hoffmann, M.; Wagner, C. S.; Harnau, L.; Wittemann, A., 3d Brownian Diffusion of Submicron-Sized Particle Clusters. *Acs Nano* **2009,** *3* (10), 3326-3334.

51. Hunter, G. L.; Edmond, K. V.; Elsesser, M. T.; Weeks, E. R., Tracking Rotational Diffusion of Colloidal Clusters. *Optics Express* **2011,** *19* (18), 17189-17202.

52. Fung, J.; Manoharan, V. N., Holographic Measurements of Anisotropic Three-Dimensional Diffusion of Colloidal Clusters. *Phys. Rev. E* **2013,** *88* (2).

53. Kraft, D. J.; Wittkowski, R.; ten Hagen, B.; Edmond, K. V.; Pine, D. J.; Löwen, H., Brownian Motion and the Hydrodynamic Friction Tensor for Colloidal Particles of Complex Shape. *Phys. Rev. E* **2013,** *88* (5), 050301.

54. Chakrabarty, A.; Konya, A.; Wang, F.; Selinger, J. V.; Sun, K.; Wei, Q.-H., Brownian Motion of Boomerang Colloidal Particles. *Physical Review Letters* **2013,** *111* (16), 160603.

55. Chakrabarty, A.; Wang, F.; Fan, C. Z.; Sun, K.; Wei, Q. H., High-Precision Tracking of Brownian Boomerang Colloidal Particles Confined in Quasi Two Dimensions. *Langmuir* **2013,** *29* (47), 14396-14402.

56. Saxton, M. J.; Jacobson, K., Single-Particle Tracking: Applications to Membrane Dynamics. *Annual Review of Biophysics and Biomolecular Structure* **1997,** *26*, 373-399.

57. Wells, N. P.; Lessard, G. A.; Goodwin, P. M.; Phipps, M. E.; Cutler, P. J.; Lidke, D. S.; Wilson, B. S.; Werner, J. H., Time-Resolved Three-Dimensional Molecular Tracking in Live Cells. *Nano Letters* **2010,** *10* (11), 4732-4737.

58. He, K.; Luo, W.; Zhang, Y.; Liu, F.; Liu, D.; Xu, L.; Qin, L.; Xiong, C.; Lu, Z.; Fang, X.; Zhang, Y., Intercellular Transportation of Quantum Dots Mediated by Membrane Nanotubes. *Acs Nano* **2010,** *4* (6), 3015-3022.

59. Wang, Z.-G.; Liu, S.-L.; Tian, Z.-Q.; Zhang, Z.-L.; Tang, H.-W.; Pang, D.-W., Myosin-Driven Intercellular Transportation of Wheat Germ Agglutinin Mediated by Membrane Nanotubes between Human Lung Cancer Cells. *Acs Nano* **2012,** *6* (11), 10033-10041.